\documentstyle[12pt,aaspp4]{article}    
\tighten    
\def\etal{et al.}

\received{1997 March 4}    
\revised{1997 April 16}    
\accepted{1997 April 18}    
\lefthead{Band}    
\righthead{GRB Crosscorrelations}    
    
\begin{document}    
    
\title{Gamma-Ray Burst Spectral Evolution Through \\
Crosscorrelations of Discriminator Light Curves}    
\author{David L. Band}    
\affil{CASS 0424, University of California, San Diego, La Jolla, CA  92093}    

\centerline{\it Received 1997 March 4; accepted 1997 April 21}
\centerline{To appear in the 1997 September 10 issue of}
\centerline{{\it The Astrophysical Journal} (Volume 486)}

\begin{abstract}
Gamma-ray burst spectra usually show hard-to-soft evolution within
intensity spikes and from spike to spike.  The techniques used to
study spectral evolution often lack sufficient temporal resolution to
determine the nature of this evolution.  By comparing the auto- and
crosscorrelations between the time histories of the BATSE Large Area
Detector discriminator rates I have characterized the spectral
evolution of a sample of 209 bursts. I find that hard-to-soft
evolution is ubiquitous, and only $\sim10$\% of the bursts show clear
soft-to-hard evolution. 
\end{abstract}
\section{Introduction}    

Gamma-ray burst spectra are not constant during a burst, but tend to soften
as the burst progresses.  Since the observed spectrum reflects the energy
content and particle distributions within the source's emitting region,
spectral variations are an important diagnostic of the nature of this
region.  Until the origin of bursts is known definitively, the physics of
the emission region will not be understood fully, and many burst phenomena
will not be explained. Indeed, theory currently provides little guidance
for the analysis of burst data. Nonetheless, the temporal and spectral
properties of bursts might reveal clues about the origin of these events,
and will provide powerful constraints on the detailed physical models which
will be constructed when the sources are identified. 

Spectral evolution has usually been demonstrated by fitting series of
spectra accumulated during the burst.  These studies (e.g., Kargatis et al.
1994; Ford et al. 1995) often lack the temporal resolution to characterize
the spectral evolution of the short timescale structure which is evident
to the eye in burst time histories.  Here I introduce and use a method
which sacrifices spectral resolution to gain the temporal resolution
necessary to determine the character of the spectral evolution of a large
burst sample.  In this study I use the terms hard and soft to refer to high
and low average photon energies, respectively. 
   
Early studies revealed two basic characteristics of spectral
evolution. By comparing the rates from two Konus detectors on {\it
Venera} 11 and 12 covering different energy ranges, Golenetskii et al.
(1984) reported that burst intensities and spectral hardness were
correlated; thus when the intensity increased, the spectrum hardened. 
On the other hand, Norris et al. (1986) found a hard-to-soft trend
across 10 bursts observed by GRS and HXRBS on the {\it Solar Maximum
Mission}. Subsequent studies showed that both trends hold in general:
the spectrum does indeed harden during intensity spikes, but there is
a hard-to-soft trend during these spikes, and the hardness tends to
peak at successively lower values from spike to spike (Kargatis et al.
1994 using SIGNE observations; Ford et al. 1995 using BATSE spectra). 
These studies lacked sufficient spectral and temporal resolution to
determine how ubiquitous these spectral evolution trends are. Most of
the earlier studies tracked the spectral evolution by fitting models
to a sequence of spectra across a number of bursts, and then comparing
the time histories of the intensity and a measure of the spectral
hardness, e.g., the energy $E_p$ of the peak of $E^2 N(E)\propto \nu
F_\nu$.  However, counts must be collected over timescales of order
seconds to accumulate a spectrum which can be fitted well; indeed,
although the BATSE Spectroscopy Detectors (SDs) can accumulate spectra
on a timescale as short as 0.128s, the temporal resolution frequently
must be degraded to get a fit which determines the spectral parameters
with relatively small uncertainties (Ford et al. 1995). However, the
time history often shows temporal structure on timescales comparable
to or shorter than the spectra's accumulation times. Thus these
studies often cannot determine the spectral trends of this short
timescale structure. 

Spectral resolution must be sacrificed to determine the spectral evolution
on the timescales of the structure evident in a burst's light curve.  In
general spectral fits are used to generate a single hardness value whose
time history is compared to that of the intensity. Thus spectral evolution
can be studied by cruder measures of spectral hardness available on shorter
timescales.  The BATSE Large Area Detectors (LADs) provide discriminator
rates every 64~ms from 2.048~s before the burst trigger until 4--10 minutes
after the trigger; background rates accumulated every 1.024~s are available
for 10 minutes before this burst data.  The discriminator rates are
available for 4 channels covering 20--50~keV, 50--100~keV, 100--300~keV and
300~keV to approximately 2~MeV.  These rates have excellent statistics
since they are collected over wide energy bands from two or three LADs with
their enormous areas of 2025 cm$^2$ each.  Spectral evolution can be
determined by comparing the light curves of each discriminator channel. 
Indeed, spectral evolution is often clearly evident to the eye; in
multispike bursts the later spikes are frequently much weaker in the
highest energy discriminator than in the lower energy discriminators. 

Throughout this paper I refer to coherent increases and decreases in the
light curve as spikes.  While such structures are apparent to the eye,
defining a spike quantitatively is complicated.  For example, Norris et al.
(1996) decompose burst light curves into pulses described by a 5 parameter
functional form. Pendleton et al. (1997) use a peak-identification
algorithm based on the depth of the minima between peaks.  Here I do not
use a rigorous definition of a spike but instead rely on a qualitative
identification.  Note that the bursts used in my analysis are strong and a
statistical fluctuation will not be mistaken for a real spike. 

The comparison of discriminator rates can reveal relative changes in
spectral hardness within a burst.  However, because the discriminators
provide only 4 channels over broad energy bands from detectors with fairly
poor spectral resolution (30\% at 88~keV), the spectra (and therefore
$E_p$) cannot be deconvolved from these rates.  Many techniques of
comparing these discriminator rates suggest themselves. For example, one
can contrast the time histories of the intensity and various ``colors,''
ratios of the discriminator rates.  Similarly, the ``orbit'' described by a
burst in the multidimensional space defined by the discriminator rates can
be studied.  Here I use the crosscorrelation functions (CCFs) between
different discriminator rates to characterize the spectral evolution of a
large burst sample (note that the autocorrelation function---ACF---is the
CCF of a rate with itself).  As will be discussed in greater detail below,
the graphical comparison of the CCFs of a fiducial channel with all the
channels can be readily interpreted.  For small lags the two time series
will have been shifted relative to each other by very little, revealing the
structure of the burst's spikes. For multispike bursts large lags will
shift a spike in one time series to another spike in the second time
series, permitting the comparison of the relative hardness of successive
spikes. 

Standard temporal analysis methods assume that the signal is
``stationary,'' that basic quantities which characterize the signal,
such as the mean value or the frequency content, remain constant
(e.g., Priestley 1981).  These techniques assume that the finite time
series under analysis is representative of the infinite time series
from which the sample has been extracted.  This is definitely not the
case for gamma-ray bursts.  These are transient events which can be
studied in their entirety. In addition, the character of the burst
changes as the burst progresses: the average intensity often
decreases, and the spectrum usually softens. ACFs and CCFs do not
assume underlying periodicities, and as I shall explain in \S 2, as
defined for transient events, the ACFs and CCFs calculated with
varying amounts of background data are the same (except for
statistical fluctuations).  On the other hand, the ACFs and CCFs are
averages over the entire burst and therefore assume that the
relationship between quantities separated by a given timescale (lag)
is the same throughout the burst.  Other techniques, such as wavelets
or pulse decomposition (Norris et al. 1996), are necessary to study
how the structure at a given timescale changes over the burst. 

ACFs and related quantities have been used before in the study of
gamma-ray bursts, primarily to study the duration of temporal
structure at different energies (which is a consequence of spectral
evolution, although this evolution was not the purpose of most of
these studies).  Link, Epstein and Priedhorsky (1993) used the ACFs of
the four LAD discriminator channels for 20 bursts to show that the
duration of temporal structures within a burst decreases for higher
energies; this was demonstrated quantitatively by integrating and
comparing the area under the ACF curves.  Their analysis did not
distinguish between the structure on different timescales.  Using the
skewness function (related to the third moment of the time history)
they also showed that bursts are asymmetric: they rise more rapidly
than they fall.  Nemiroff et al. (1994) confirmed that bursts are
asymmetric.  Fenimore et al. (1995) averaged the ACFs in each of the
LAD discriminator channels of 45 bright bursts to find the average ACF
as a function of energy. They found that the width of the temporal
structure decreases with increasing energy as a power law $E^{-0.4}$,
and that the average ACF can be parameterized as the sum of two
exponentials.  In't Zand and Fenimore (1996) analyzed the properties
of burst ACFs in terms of a modified shot noise model, and concluded
that there are no hidden systematic trends which preclude the use of
ACFs to study the duration of temporal structure (e.g., possible
cosmological time dilation). Kouveliotou et al. (1992) used a Fourier
cross-spectral technique (using quantities which are Fourier
transforms of the CCF) to show that the hard channels lead the soft
channels in 90\% of 22 bursts. 

The calculation of CCFs for transient events involves some subtleties;
the methodology is developed in \S 2.  The interpretation of the
resulting CCFs is explored in \S 3.  Then I present the results of
applying this technique to a large burst sample (\S 4).  Finally, I
discuss the implications of these results (\S 5). 
\section{Methodology}    
The CCF measures the temporal correlation of two time series $v_1$ and
$v_2$: 
\begin{equation}    
CCF(\tau;v_1 , v_2) = {{\langle v_1(t) v_2(t+\tau)\rangle}\over    
   {\sigma_{v1} \sigma_{v2}}} \quad \hbox{ where } \quad   
   \sigma_v = \sqrt{\langle v^2 \rangle} \quad ,
\end{equation}    
and $\tau$ is called the ``lag.'' If $v_2$ leads $v_1$ by $t_0$ (e.g.,
$v_1(t)=v_2(t-t_0)$) then the CCF peaks at $\tau=-t_0$.  The ACF is
the CCF of a time series with itself: 
\begin{equation}    
ACF(\tau ;v) = CCF(\tau; v,v) \quad .    
\end{equation}    
Note that the ACF is symmetrical, and that $ACF(\tau=0;v)=1$.      

In most applications, the CCF is calculated for an infinitely long
function $d(t)$ which is sampled over a short time range; the sample
is assumed to be characteristic of all time ranges (i.e., the signal
is assumed to be ``stationary'').  Since the interest is in the
varying component of the signal, the signal is regarded as
fluctuations around a constant value, and the CCF is calculated using
a zero-mean time series: 
\begin{equation}    
v(t) = d(t) - \langle d(t) \rangle    
\end{equation}    
However, a gamma-ray burst is a transient phenomenon which is
contained entirely within the sampled time range.  If the transient
event were treated as a stationary event, the background rate before
and after the burst would have a negative value in the zero-mean time
series sampled over a finite sampling window and the calculated CCF
would depend on the size of this window. A burst time series acquired
with an infinite data window would have a mean equal to the background
rate (if constant) since the counts in the burst would be diluted by
the infinite time range, and the background in the zero-mean time
series would fluctuate around zero. Therefore, I define the CCF using
a background-subtracted time series 
\begin{equation}    
v(t) = d(t) - b(t)    
\end{equation}    
which emphasizes the signal of interest (Link et al. 1993; Fenimore et
al. 1995).  Using this definition of $v(t)$ the difference between the
CCFs calculated for the infinite and finite data windows should be
small since the background omitted in the finite data window would
have contributed nothing, on average, to all averages. Thus the
effective data window can be considered to be greater than the time
range actually sampled. 
    
The observed signal $v(t)$ includes noise $n(t)$. We would like to
find the ACF and CCF for the signal alone, mitigating the bias
introduced by the presence of noise.  I assume the noise has the
following properties: 
\begin{equation}    
\langle n \rangle = 0  \quad \hbox{and} \quad  
\langle n(t)n(t+\tau) \rangle = \sigma_n^2 \delta(\tau) \quad .   
\end{equation}    
Note that $\langle n(t)n(t+\tau) \rangle$ is truly 0 for $\tau\ne 0$
only for an infinite time series (or only on average), and I expect
small fluctuations around 0 for any particular finite time series. 
The background-subtracted time series is the sum of the signal $s(t)$
and the noise $n(t)$, 
\begin{equation}   
v(t) = s(t)+n(t) \, , \quad \hbox{and} \quad    
\sigma_v^2 = \sigma_s^2 +\sigma_n^2 +2\langle sn\rangle \quad .   
\end{equation}    
I do not consider the uncertainty in the determination of the
background $b(t)$ which introduces an error which might violate both
properties in eq.~(5). The ACF calculated with $\sigma_v$ is 
\begin{equation}    
ACF(\tau; v) = ACF(\tau; s) {{\sigma_s^2}\over{\sigma_v^2}}    
   + {{\sigma_n^2}\over{\sigma_v^2}}\delta(\tau)    
   +{{\langle n(t) s(t+\tau) + s(t) n(t+\tau) \rangle}\over{\sigma_v^2}}    
\end{equation}    
while the CCF is 
\begin{eqnarray}    
CCF(\tau; v_1 , v_2) &=& CCF(\tau; s_1,s_2)    
   {{\sigma_{s1}\sigma_{s2}}\over{\sigma_{v1}\sigma_{v1}}} \\    
&+& {{\langle n_1(t) n_2(t+\tau) \rangle}\over{\sigma_{v1}\sigma_{v2}}}    
   +{{\langle n_1(t) s_2(t+\tau) + s_1(t) n_2(t+\tau) \rangle}    
   \over{\sigma_{v1}\sigma_{v2}}} \quad . \nonumber    
\end{eqnarray}    
On average I expect $\langle n_1(t) n_2(t+\tau) \rangle=0$ and
$\langle s(t) n(t+\tau) \rangle=0$.  If the magnitude of the noise $n$
is a function of the signal $s$ (e.g., $n$ results from Poisson
statistics for binned data), the average $\langle ns \rangle$ should
be close to 0 because $n$ has zero mean, although these averages will
actually fluctuate around 0 because of the finite number of data
points. Note that the noise reduces the signal's ACF and CCF, and
produces a spike at zero lag in the ACF. To correct for this effect of
noise, I change the variance used in the denominators of the ACF and
CCF to 
\begin{equation}    
\sigma_v^{\prime\,2} = \sigma_v^{2}-\sigma_n^2 
\end{equation}    
(Link et al. 1993; Fenimore et al. 1995). For noisy signals this
expression can underestimate the variance, and thus overestimate the
CCF and ACF. 

Since the time series is sampled at a number of discrete data points,
averages are calculated by sums, not integrals. Assume there are $N$
data points which correspond to the counts accumulated during time
bins of duration $\Delta t=T/N$; the time series spans a time range
$T$.  I define the quantities as follows: 
\begin{eqnarray}    
v_i &=& d_i-b_i \quad , \nonumber \\    
\sigma_v^{\prime \, 2} &=& {1\over N} \sum_{i=1}^N (v_i^2 - d_i)  \quad , \\    
ACF(\tau=k\Delta t; v) &=& {{\sum_{i=max(1,1-k)}^{min(N,N-k)} v_i v_{i+k}}     
   \over {N\sigma_v^{\prime \,2}}} \quad , \nonumber \\    
CCF(\tau=k\Delta t; v_1,v_2) &=& {{\sum_{i=max(1,1-k)}^{min(N,N-k)}     
   v_{1i} v_{2(i+k)}} \over {N\sqrt{\sigma_{v1}^\prime\sigma_{v2}^\prime}}}     
   \quad . \nonumber    
\end{eqnarray}    
For these discretized equations $d_i$ is the observed data point from
the $i$th time bin, $b_i$ is the background rate (usually interpolated
from data before and after the burst), and $v_i$ is the
background-subtracted signal.  If there are more than one time series,
then $v_{1i}$ is the $i$th data point of the first time series, etc. I
have assumed that the noise results from Poisson fluctuations, and
therefore $\sigma^2_n=\langle d_i \rangle$. In the ACF and CCF there
are $N-|k|$ products in each sum for a lag of $\tau=k\Delta t$, yet
they are divided by $N$ to form the average. As discussed above, I am
effectively assuming that the products in the time range where one
time series extends beyond the other are approximately zero, since at
least one of the time series will be background over that time range. 
In the absence of noise the ACF and CCF will not change with
variations in the amount of background data included before and after
the burst. Similarly, $\sigma_v^{\prime\, 2}$ should be independent of
how much background is included since on average $v_i^2 \sim d_i$ for
background data points.  Yet $\sigma_v^{\prime\, 2}$ will be sensitive
to fluctuations and small errors in the background determination, and
therefore the most accurate CCFs and ACFs are obtained by including
only the burst within the data window. 

A measure of the uncertainty is desirable for any observed or
calculated quantity.  The uncertainty on the ACFs and CCFs can be
estimated by propagating the uncertainty on each element of the two
time series (see also Fenimore et al. 1995). This calculation requires
the partial derivatives of the ACF and CCF with respect to each time
series element, an algebraically complicated quantity not presented
here.  The uncertainties in the ACF and CCF values at different lags
are correlated both because the uncertainty in the normalization is a
component of the overall uncertainty and because each ACF and CCF
value is a function of each time series element.  In general the
uncertainties are small. 

To study the effect of noise empirically, I calculated the ACF and
CCFs for time series with different signal-to-noise ratios (SNRs), and
found significant deviations only for low SNRs.  For example, I
computed the CCF for a $\sim1$~s pulse where one time series was very
strong and the second one was weak.  I find significant deviations
(the rising and falling segments of the CCF differ noticeably from the
asymptotic value) when the peak time bin of the weak time series had
an SNR less than 3 (the background count rate was 3000~cts~s$^{-1}$,
the time bins were 0.064~s wide and the signal/background at the peak
was 0.24).  Only for much smaller SNRs do the correlation curves
become noisy. In practice the signal in the highest energy
discriminator (300--2000~keV) was often very weak or nonexistent and
the CCF with this discriminator channel was very noisy.  In such cases
I neglected the CCF with this channel in the analysis. 

The discriminator time series are constructed from three different LAD
data types---DISCLA, PREB and DISCSC (Fishman et al. 1989).  The
result has 1.024~s resolution up to 2.048~s before the burst trigger,
and 0.064~s thereafter.  The CCFs and ACFs are calculated with the
minimum data necessary to include the burst. Occasionally the burst
began more than 2.048~s before the trigger when the time bins were
still 1.024~s wide rather than 0.064~s.  Since I want to include all
the identifiable burst emission, and the ACF and CCF require data with
uniform time bins, I replace each of these 1.024~s bins with 16
0.064~s bins with the same count rate.  To insure that these new
0.064~s bins have the appropriate noise characteristics, I add
Gaussian fluctuations. In general the background is fitted with a
quadratic polynomial of time using background data before and after
the burst.  I calculate and plot the ACF for discriminator 3
(100--300~keV) and the CCFs for discriminators 1 (20--50~keV), 2
(50--100~keV) and 4 (300--$\sim2000$~keV) with discriminator 3, i.e.,
$ACF(\tau;v_3)$ and $CCF(\tau;v_3, v_1)$, $CCF(\tau;v_3, v_2)$, and
$CCF(\tau;v_3, v_4)$. These four functions are then compared both
graphically and quantitatively. 
\section{Interpretation of ACFs and CCFs}   
For every burst in my sample I calculate and plot the ACF for
discriminator 3 and the CCFs for discriminators 1, 2 and 4 with
discriminator 3.  Thus the spectral evolution must be characterized
from the differences between the ACF and the CCFs.  Here I present a
number of heuristic examples which reveal the features which can be
extracted from the graphical comparison of these curves. In these
examples I compare two time series, $v_h$ and $v_s$, meant to
represent high and low energy channels, respectively, and I calculate
$ACF(\tau; v_h)$ and $CCF(\tau ; v_h , v_s)$. 

First, consider $v_h = \exp(-t/t_h)$ and $v_s = \exp(-t/t_s)$ for $t
\ge 0$. If $v_h$ and $v_s$ are the hard and soft channels,
respectively, then $t_s > t_h$ for hard-to-soft evolution.
Consequently 
\begin{eqnarray}
ACF(\tau ; v_h) &=& \exp(-|\tau|/t_h) \quad , \\
CCF(\tau ; v_h , v_s) &=& \cases{2 {\sqrt{t_h t_s}\over {t_h+t_s}} 
   \exp(-\tau/t_s) \quad , & $\tau > 0$ \quad, \cr
   2 {\sqrt{t_h t_s}\over {t_h+t_s}} \exp(\tau/t_h) 
   \quad , & $\tau \le 0$ \quad . \cr}
\end{eqnarray}
Thus for hard-to-soft evolution of falling exponential spikes the CCF
and ACF will be parallel on the negative lag side, while the CCF will
decrease more slowly on the positive lag side.  If $t_s$ is not much
larger than $t_h$ (as is often the case), then the maximum CCF value
is not much less than 1. Consequently the ACF and CCF will nearly
coincide on the negative lag side, and the CCF will exceed the ACF on
the positive lag side (see Figure~1a). 

Now consider rising exponentials.  In this case the ACF is the same as
above, but the CCF is time reversed:  $CCF \propto \exp(-t/t_h)$ for
$\tau > 0$ and $CCF \propto \exp(t/t_s)$ for $\tau \le 0$.  However,
now $t_s < t_h$ for hard-to-soft evolution:  the CCF rises more
rapidly than the ACF on the negative lag side, and is parallel to the
ACF on the positive side (see Figure~1b). 

These examples with exponential spikes demonstrate two features of the
comparison between ACFs and CCFs.  First, if the ACF is calculated for
the hard channel, then hard-to-soft evolution is indicated by the CCF
rising more rapidly on the negative lag side or falling more slowly on
the positive lag side than the ACF.  Second, the relative widths of
the ACF and CCF indicate the relative widths of the spike in the hard
and soft channels. 

However, the width of an ACF or CCF peak must be interpreted with
care. Here I define the width to be the FWHM of the peak in the time
series or ACF.  The relation between the width in the time series and
the ACF depends on the shape of the peak.  For a square pulse of width
$\Delta t$ the $ACF=1-|\tau|/\Delta t$, and thus the signal and ACF
widths are the same. For a triangular pulse (infinitesimal rise time,
linear decay) of length $\Delta t$ the $ACF = (1-|\tau|/\Delta
t)^2(1+|\tau|/2\Delta t)$, and the ACF width is $\sim 4/3$ the pulse
width.  For a Gaussian pulse $v\propto \exp[-(t/\sigma)^2]$ the
$ACF=\exp[-t^2/2\sigma^2]$, and thus the ACF width is $\sqrt{2}$ times
the pulse width.  Finally, as shown above, an exponential pulse has an
ACF with a width twice that of the signal.  The actual profiles of
spikes in bursts are complicated, and therefore the widths of the
ACFs, and by analogy the CCFs, cannot be translated quantitatively
into measures of the spike widths. 

To interpret well-separated secondary peaks in the ACF and CCF when
the lag is great enough so that different spikes coincide, consider a
burst which consists of two spikes of unit width.  Without loss of
generality the height of the first peak in each time series can be set
to unity.  Then $v_h(t) = \delta(t)+ a_h \delta( t - \Delta t)$ and
$v_s(t) = \delta(t)+a_s \delta(t-\Delta t)$, where $\Delta t$ is the
separation between peaks. Hard-to-soft evolution from spike to spike
results from $a_s > a_h$; it does not matter whether $a_s$ or $a_h$
are greater than 1.  The ACF of $v_h$ is 
\begin{equation}  
ACF(\tau; v_h) = \delta(\tau)+{{a_h}\over{1+a_h^2}} \delta 
   (\tau \pm \Delta t)  
\end{equation}  
and the CCF of $v_h$ and $v_s$ is  
\begin{equation}  
CCF(\tau; v_h,v_s) = {{1+a_h a_s}\over\sqrt{(1+a_h^2)(1+a_s^2)}} 
   \delta(\tau)+  
   {{a_h}\over{1+a_h^2}} \sqrt{{1+a_h^2}\over{1+a_s^2}}   
   \left( \delta (\tau + \Delta t) + {{a_s}\over{a_h}} 
   \delta (\tau-\Delta t) \right) \quad .  
\end{equation}  
If $a_s>a_h$ then the CCF will be greater than the ACF at $\tau =
\Delta t$. The comparative heights of the secondary peaks in the ACF
and CCF on the positive lag side indicates the relative magnitudes of
$a_s$ and $a_h$, and thus the nature of the spike-to-spike spectral
evolution. 

I conclude that for the ACFs and CCFs calculated from the BATSE data,
hard-to-soft evolution is indicated by $CCF(\tau; v_3,v_4)<$
$ACF(\tau; v_3)<$ $CCF(\tau; v_3,v_2)<$ $CCF(\tau; v_3,v_1)$ on the
positive lag side or the opposite on the negative lag side. 
\section{Results}  
My burst sample consists of 209 of the strongest BATSE bursts from the
beginning of the mission through 15~August 1996.  In general, the
bursts had a peak count rate in the 50--300~keV band greater than
10,000~cts~s$^{-1}$, but some large fluence bursts with a smaller peak
count rate were included. Almost all the bursts I selected were longer
than 1~s so that there would be enough points to produce ACFs and CCFs
with structure. Consequently this burst sample is not statistically
complete. Because of the preference for long, strong bursts, many of
the bursts consisted of several clusters of emissions separated by
periods of little or no emission above background. 

Figure 2 shows six bursts:  the upper panel is the time history
summing all the discriminator channels, and the bottom panel is a
comparison of the $CCF(\tau; v_3,v_4)$ (3~dots-dashed curve),
$ACF(\tau; v_3)$ (solid curve), $CCF(\tau; v_3,v_2)$ (dot-dashed
curve), and $CCF(\tau; v_3,v_1)$ (dashed curve). The order and
relative widths of $CCF(\tau; v_3,v_4)$, $ACF(\tau; v_3)$, $CCF(\tau;
v_3,v_2)$, and $CCF(\tau; v_3,v_1)$ on the positive lag side of bursts
3B~910503, 3B~911118, and 3B~920216 show that these bursts underwent
hard-to-soft evolution (bursts are identified by their names in the 3B
Catalog of Meegan et al. (1996), except for bursts which occurred
after this catalog). In particular, $CCF(\tau; v_3,v_4)<$ $ACF(\tau;
v_3)<$ $CCF(\tau; v_3,v_2)<$ $CCF(\tau; v_3,v_1)$ in the secondary
structure at $\tau=40$--50~s of the correlations for 3B~910503
indicate that the first emission cluster of this burst at $t=0$--10~s
was harder than the second cluster at $t=46$--54~s. On the other hand
3B~920701 and 3B~930612B are among the very rare cases of soft-to-hard
evolution; note that the order of the correlations is reversed
compared to the usual hard-to-soft evolution. Finally, GB~941119 is an
example of a burst without clear evolution. 

Emission spikes or clusters of spikes which are well-separated (e.g.,
separations many times their width) result in correlations with
structure that can clearly be identified as the correlation of a
cluster with itself (around zero lag) or with other clusters, as is
the case for 3B~910503.  On the other hand, the correlations may
smooth over the structure within a cluster of closely spaced spikes
(e.g., spacing of order the width of a spike); again, the central peak
of the correlations for 3B~910503 is such an example. While there may
be clear short duration emission spikes ($\sim$1s long), the central
ACF and CCF peaks are often much broader, as is the case for
3B~9110503 and GB~941119. Thus, if many emission spikes occur in a
cluster, then the width of the ACFs' and CCFs' central peak may
reflect the duration of the cluster, and not of the individual spikes.

Hard-to-soft spectral evolution is found in most bursts. I categorized
the spectral evolution evident in the central and secondary peaks in
the ACFs and CCFs.  To those bursts which show unambiguous
hard-to-soft or soft-to-hard evolution I assigned a value of +2 or -2,
respectively.  Note that the evolution need not be extreme for a value
of $\pm2$.  Where the signature for spectral evolution was somewhat
ambiguous I assigned $\pm1$. Finally, I gave a 0 to those bursts with
no indication of spectral evolution.  Thus positive and negative
values indicate hard-to-soft and soft-to-hard evolution, respectively,
and the magnitude denotes the confidence with which I could determine
the trend.  I assigned values for both the central peak and the
secondary structure. The results of this analysis are tabulated in
Table~1; since there are just over 200 bursts in my sample, each burst
corresponds to approximately 0.5\% of the sample. In 9 cases the
secondary structure showed both types of evolution, in which case I
split the burst between the two types of evolution. If the burst was
very simple and there was no secondary structure, I assigned to the
secondary structure the same value as the central peak. As is clear
from the table, hard-to-soft spectral evolution is a standard
characteristic of gamma-ray bursts.  However, there are bursts with
soft-to-hard evolution, although only $\sim2$\% show such evolution in
both the central peak and the secondary structure.  Soft-to-hard
evolution is more common in the secondary structure than in the
central peak. 

The evolutionary trend in the central peak can be determined
quantitatively by considering the time lag of the peaks of the CCFs,
as are shown by the cumulative distributions in Figure~3.  These lags
are related to the time shifts found by Kouveliotou et al. (1992)
using cross Fourier transforms (which are the transforms of CCFs). As
is clear, the CCFs of the higher energy channels peak before those of
lower energies, typically by $\sim0.1$~s, indicating that the light
curve at higher energies peaks earlier than at lower energies.
Similarly, the CCFs for the lower energy channels are broader than for
higher energies. Figure~4 shows the cumulative distribution of the
ratios of the CCFs and the ACF at a correlation value of 0.5; the
differences in the width are typically $\sim20$\%.. 

Attempts to characterize burst morphology have generally been
unsuccessful. One apparent class consists of FREDs---Fast Rise,
Exponential Decays---whose name describes them.  My database includes
35 FREDs or similarly simple bursts, nine bursts which appear to
consist of two or more FREDs and one case of an ``inverse FRED.'' 
With one exception (GB~930612B, included in Figure~2) which has more
structure than the standard FRED, all these bursts show unambiguous
hard-to-soft evolution, although not necessarily between FREDs in a
multi-FRED burst.  Similarly, Bhat et al. (1994) found that
hard-to-soft evolution predominates in FREDs; they also report that
the high energies lead the low energies by a time proportional to the
FRED's rise time.  Note that Norris et al. (1996) found a similar
trend between pulses resulting from deconvolving the LAD light curves.

Many bursts have well separated ``clumps'' of emission, and in many
cases the character of these clumps differs within a burst:  some
consist of spikes which barely overlap, whereas others have a few
spikes which protrude from a smooth envelope.  This type of
characterization remains to be quantified. By comparing the ACFs and
CCFs at both positive and negative lags I find that in some bursts one
emission clump shows very marked spectral evolution whereas a second
shows very little.  Although these phenomena need to be explored
further, they suggest that there are qualitatively different emission
processes which can occur within the same burst. 
\section{Discussion}
My analysis of the ACFs and CCFs of a sample of strong bursts shows
that within individual spikes or clusters of spikes, $\sim90$\% of the
bursts show hard-to-soft evolution and in only $\sim3$\% is the
opposite trend evident. Similarly, hard-to-soft evolution from spike
to spike or among successive clusters of spikes characterizes $\sim
80$\% of the bursts, while $\sim15$\% show the opposite trend.  Thus
hard-to-soft evolution is a very common feature of gamma-ray bursts,
but counterexamples exist.  This feature of burst phenomenology must
ultimately be explained by any successful burst model. 

For example, in the current fireball models a relativistically
expanding fireball radiates at the shocks formed either as the ejecta
within the fireball plows into the surrounding medium (Rees \&
M\'esz\'aros 1992; M\'esz\'aros \& Rees 1993; M\'esz\'aros, Rees \&
Papathanssiou 1994; Katz 1994; Sari, Narayan \& Piran 1996) or as a
consequence of inhomogeneities within the expanding fireball (Rees \&
M\'esz\'aros 1994; Paczy\'nski \& Xu 1994; Papathanssiou \&
M\'esz\'aros 1996).  At an external shock multiple spikes result from
energy release events which cool on a timescale short compared to the
expansion (Sari et al. 1996).  The Lorentz factor of the external
shock should drop as the fireball expands and the external medium is
swept up. Spike-to-spike spectral evolution can be explained easily by
later energy release events occurring at larger fireball radii.
However, internal shock models attribute multiple, well-separated
spikes to the shocks which form at different times and radii as
inhomogeneities in the flow within the fireball cause different
regions to collide; for such models it is not obvious why there should
be a hard-to-soft trend from spike to spike.  Whether the electron
acceleration event occurs at an internal or external shock, the
spectral evolution resulting from the electron cooling must be
reconciled with the observations. 

Detailed physical implications of spectral evolution are model
dependent (e.g., the above discussion of the fireball scenarios) but
some general conclusions are possible.  That the average photon energy
increases when the photon flux increases suggests that energy has been
injected into the emission region, increasing the average energy of
the emitting particles. The hard-to-soft trend from spike to spike
indicates that if there is one emission region, it retains a memory of
previous emission events. If there are multiple regions, they may
communicate, or alternatively, independent regions may possess similar
internal ``clocks'' such that they emit softer spikes later in the
burst. 

The ACF and CCF methodology can be extended to address deficiencies in
the current study. First, the ACFs and CCFs assume that the
correlation between quantities with a given time separation is the
same throughout the burst. Therefore the ACFs and CCFs of individual
spikes and clusters of spikes within a burst can be calculated
separately to characterize their spectral evolution. Structure in
different parts of a burst frequently appears very different.  For
example, 3B~910503 (see Figure~2) begins with a cluster of spikes
while the emission which begins 45~s after the burst trigger is much
smoother and simpler; thus the ACFs and CCFs of the two emission
clusters can be compared. Presumably the temporal and spectral
structure reflects the processes by which the emission region is
energized and radiates. 

Second, when intensity spikes occur in clusters where the separation
between spikes is comparable to the width of the spikes, the effect of
the spikes' structure cannot be separated from that of the overall
cluster. Filtering the time history (e.g., using an FFT filter or
wavelets) before calculating the ACFs and CCFs may separate the
structure on different timescales. This combination of filtering and
CCFs has been applied to solar flares (e.g., Aschwanden \etal\ 1996). 

Third, I restricted myself to bursts with durations over 1~s because
of the 0.064~s time resolution of the LAD light curves I used. Greater
time resolution can be obtained from the TTE data type (which provides
the time necessary to accumulate a given number of counts) for bright
bursts.  Although it would have to be modified (the TTE rates are not
provided on a uniform grid), the methodology presented here can be
applied to the TTE data to study the spectral evolution of short
bursts. 
\section{Summary}  
Spectral evolution can be characterized by comparing the
crosscorrelations of a fiducial burst intensity light curve with the
light curves in different energy bands.  The structure of the
crosscorrelations in the peak around zero lag results from the
spectral evolution of individual intensity spikes or clusters of
spikes.  Secondary peaks in the crosscorrelations away from zero lag
result from well-separated spikes or clusters of spikes. Thus the
structure of the crosscorrelations at these secondary peaks
characterize the peak-to-peak spectral evolution. 

I applied this crosscorrelation technique to a sample of 209 bursts
observed by BATSE using rates from the 4 LAD discriminator channels. 
My fiducial light curve was the count rate in the 100--300~keV band. 
I find hard-to-soft evolution is common, although there are a few
counterexamples. The preference for hard-to-soft evolution is somewhat
stronger for individual spikes and clusters of spikes than for
well-separated structure. The shift in the centers of the peaks around
zero lag between the crosscorrelations with low energy (20--50~keV)
and high energy (300--2000~keV) channels is of order 0.1~s. 
\acknowledgments    
I thank my collaborators on the BATSE team for their assistance over
the past few years. In addition, I am grateful for the careful reading
of this paper by the referee, J.~Norris, which has improved the text's
clarity. BATSE research at UCSD is supported by NASA contract
NAS8-36081. 
    
%

\clearpage    

\figcaption{$ACF(\tau;v_h)$ (solid curve) and $CCF(\tau;v_h,v_s)$ (dashed
curve) for exponential spikes.  In panel (a) the spikes are falling
exponentials $v_{h,s}\propto \exp(-t/t_{h,s})$ for $t\ge 0$ with $t_h=1$
and $t_s=1.5$, while in panel (b) the spikes are rising exponentials
$v_{h,s}\propto \exp(t/t_{h,s})$ for $t\le 0$ with $t_h=1$ and $t_s=2/3$. 
Thus both cases are characterized by hard-to-soft spectral evolution.} 

\figcaption{Intensity time history (top panel) and correlation functions
(bottom panel) for 6 bursts.  The intensity is summed over the 4 LAD
discriminators, covering 20--2000~keV.  Extending over the time range
included in the correlation functions, the (nearly) horizontal line in the
time history plot is the calculated background rate.  The correlation
functions are $CCF(\tau;v_3,v_4)$ (the CCF of LAD discriminator 3 with
discriminator 4---3 dots-dashed curve), $ACF(\tau;v_3)$ (the ACF of
discriminator 3---solid curve), $CCF(\tau;v_3,v_2)$ (discriminator 3 with
discriminator 2---dot-dashed curve), and $CCF(\tau;v_3,v_1)$ (discriminator
3 with discriminator 1---dashed curve). In general, hard-to-soft evolution
is indicated by $CCF(\tau;v_3,v_4)<$ $ACF(\tau;v_3)<$ $CCF(\tau;v_3,v_2)<$
$CCF(\tau;v_3,v_1)$ on the positive lag side.} 

\figcaption{Cumulative distributions of the lags of the center of the
central peak.  The curves are for $CCF(\tau;v_3,v_4)$ (3 dots-dashed
curve), $ACF(\tau;v_3)$ (solid curve---the ACF defines zero lag),
$CCF(\tau;v_3,v_2)$ (dashed curve), and $CCF(\tau;v_3,v_1)$ (dot-dashed
curve).  The plot shows that on average the higher energy light curves peak
earlier than the lower energy light curves.} 

\figcaption{Cumulative distributions of the ratio of the CCF to ACF widths
at a correlation value of 0.5.  The curves are for
$CCF(\tau;v_3,v_4)/ACF(\tau;v_3)$ (3 dots-dashed curve),
$CCF(\tau;v_3,v_2)/ACF(\tau;v_3)$ (dashed curve), and
$CCF(\tau;v_3,v_1)/ACF(\tau;v_3)$ (dot-dashed curve).  The plot shows that
on average the lower energy light curves are broader than the higher energy
light curves.} 

\clearpage
  
\begin{deluxetable}{l r c c c c c c r}
\tablecolumns{9}
\tablewidth{0pc}
\tablecaption{Spectral Evolution of the Central Peak and 
the Secondary Structure}
\tablehead{
%
%
&&& \multicolumn{5}{c}{Central Peak} & \\
&&\qquad \qquad & \colhead{-2} & \colhead{-1} & \colhead{0} & 
\colhead{+1} & \colhead{+2} & \qquad Total}
\startdata
&+2 && 2 & 1 & 8 & 16.5 & 121   & 148.5           \nl 
Secondary
&+1 && 0 & 0 & 0 &  8   &   8   &  16\phantom{.5} \nl 
Structure
&0  && 0 & 0 & 1 &  6   &   9   &  16\phantom{.5} \nl 
&-1 && 0 & 1 & 2 &  2   &   5.5 &  10.5           \nl 
&-2 && 3 & 0 & 2 &  2.5 &  10.5 &  18\phantom{.5} \nl
\hline
Total &&& 5 & 2 & 13 & 35 & 154 & 209
\enddata
\tablecomments{Positive and negative values indicate hard-to-soft and
soft-to-hard evolution, respectively.  A magnitude of 2 was assigned when
the evolution was unambiguous, while a magnitude of 1 indicates ambiguity. 
The absence of apparent evolution resulted in value of 0.  If the secondary
structure showed both types of evolution, half the burst was assigned to
each category. Note that the value does not indicate whether the spectral
evolution is extreme.} 
\end{deluxetable}

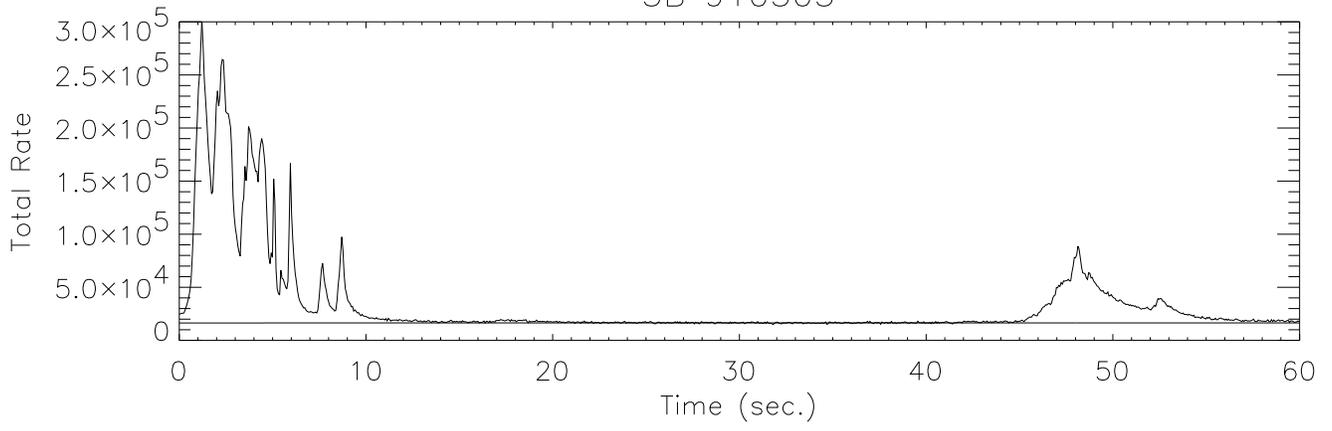
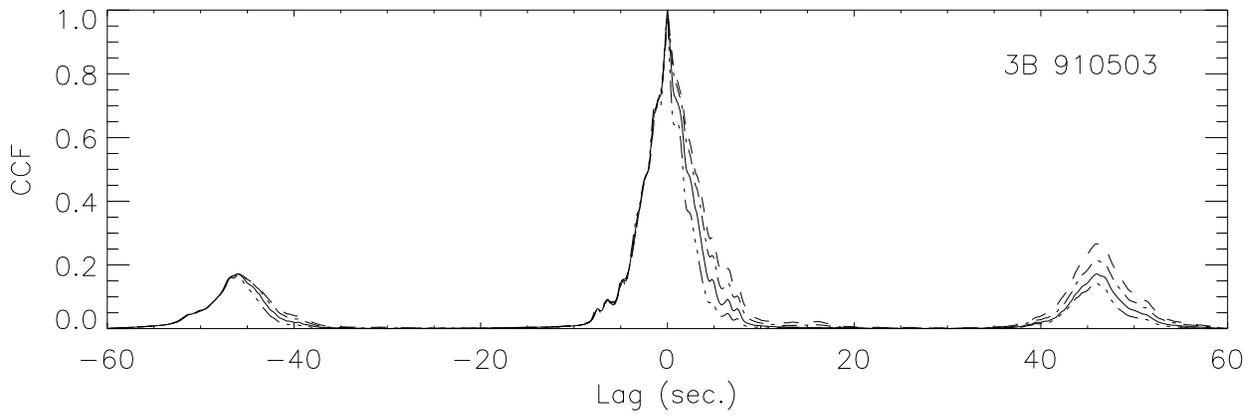
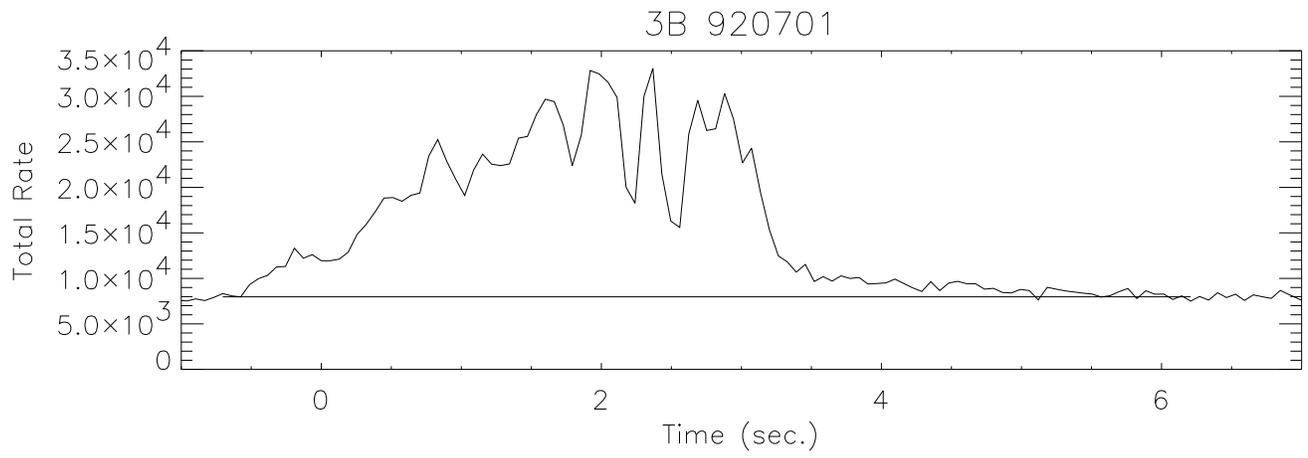
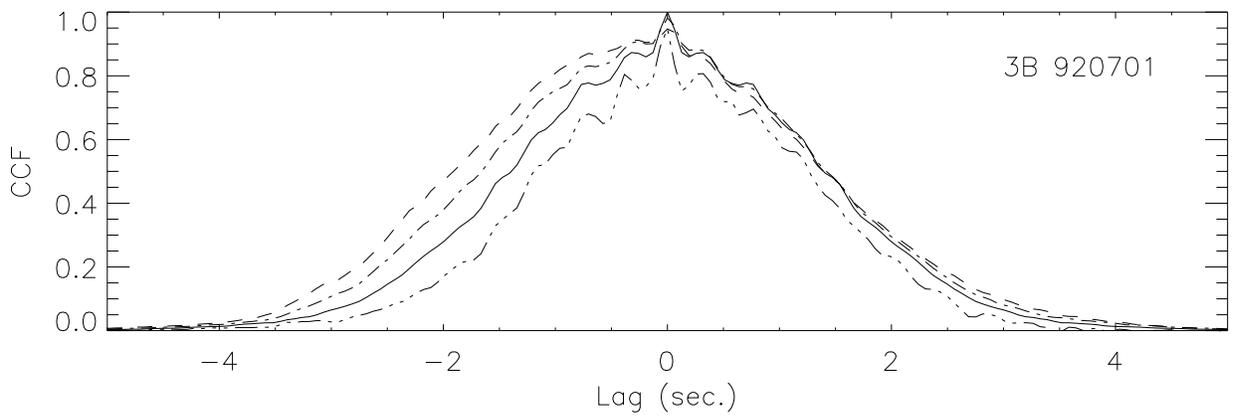
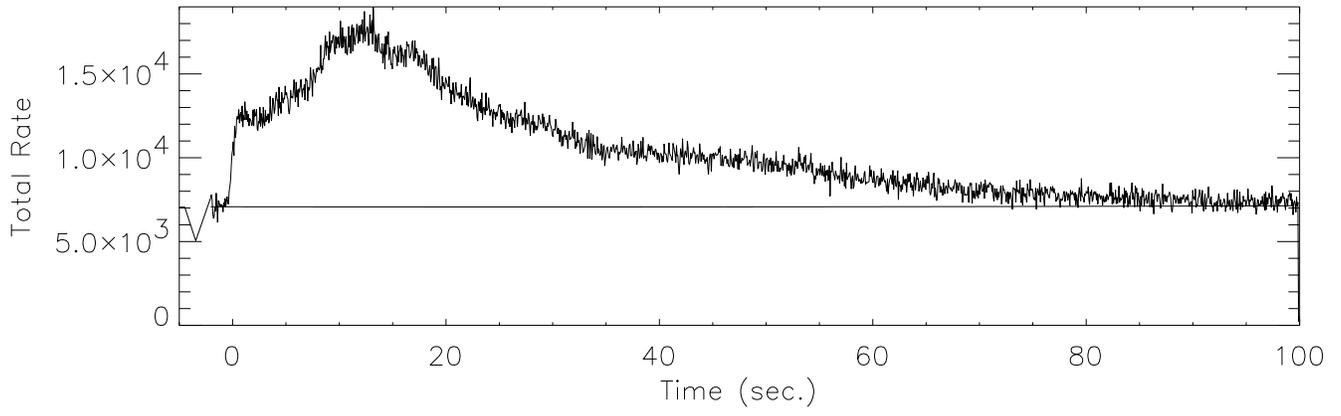
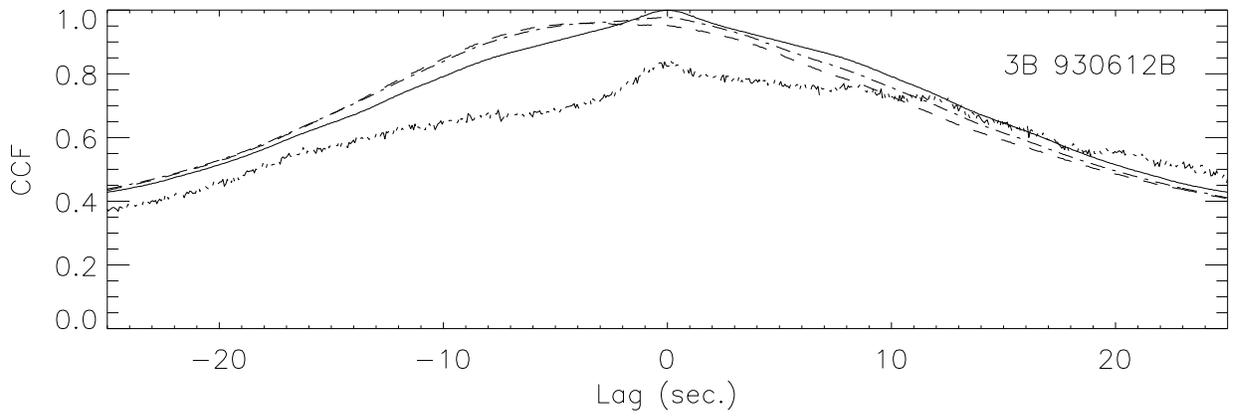
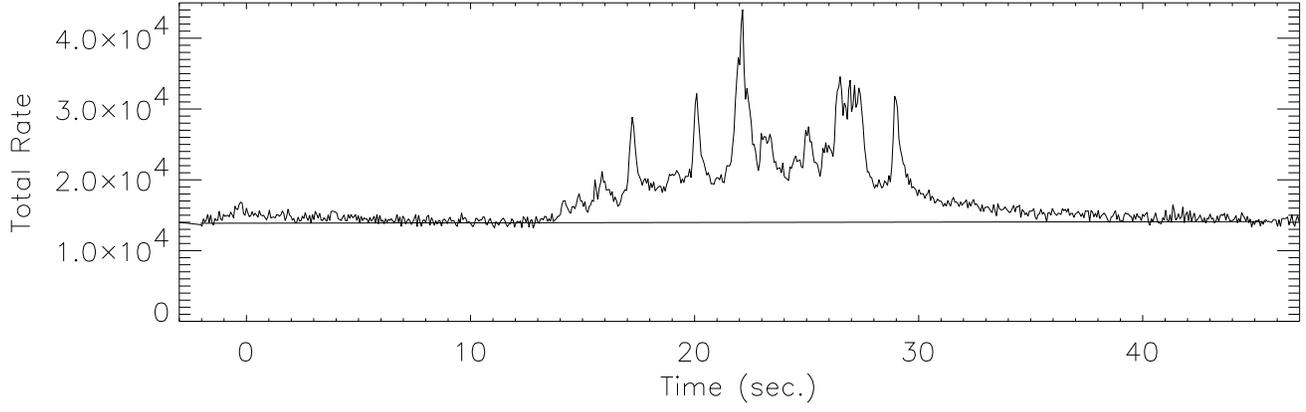
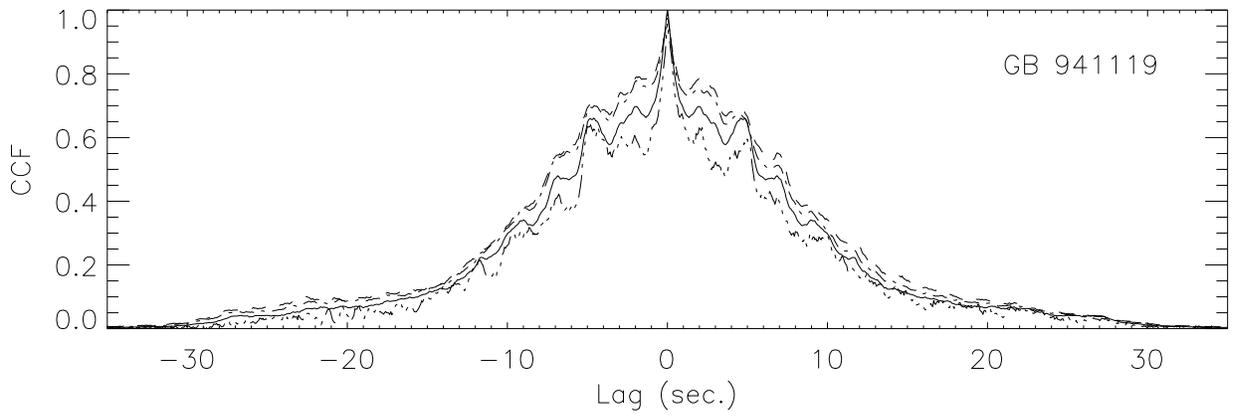
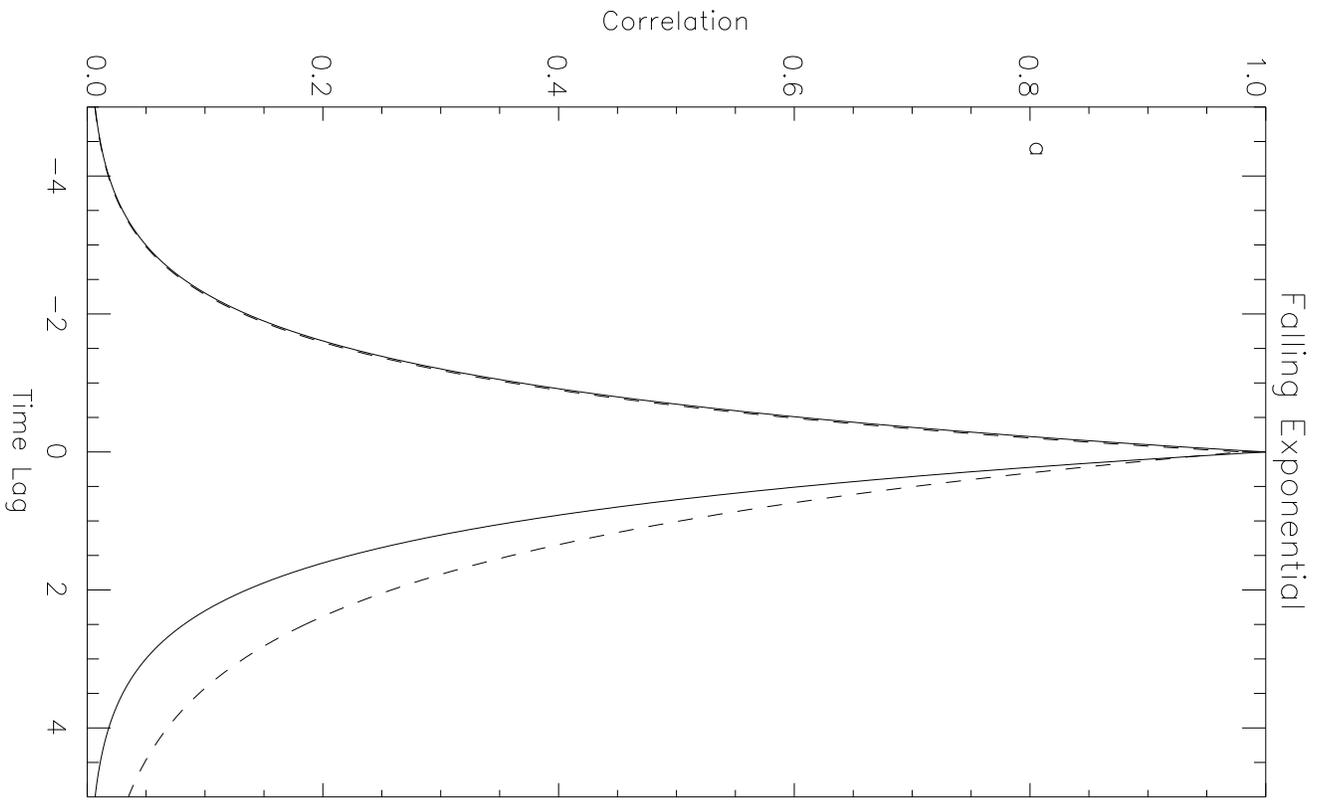
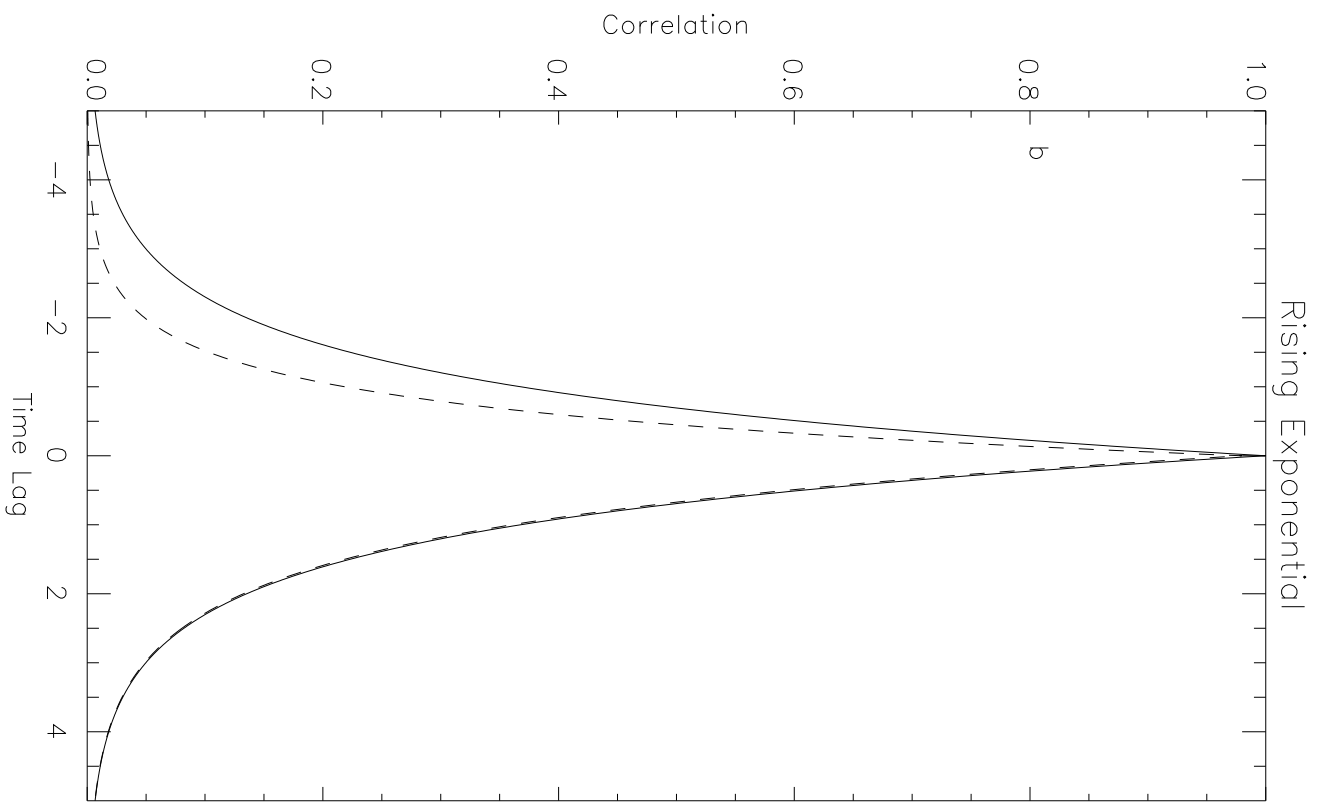
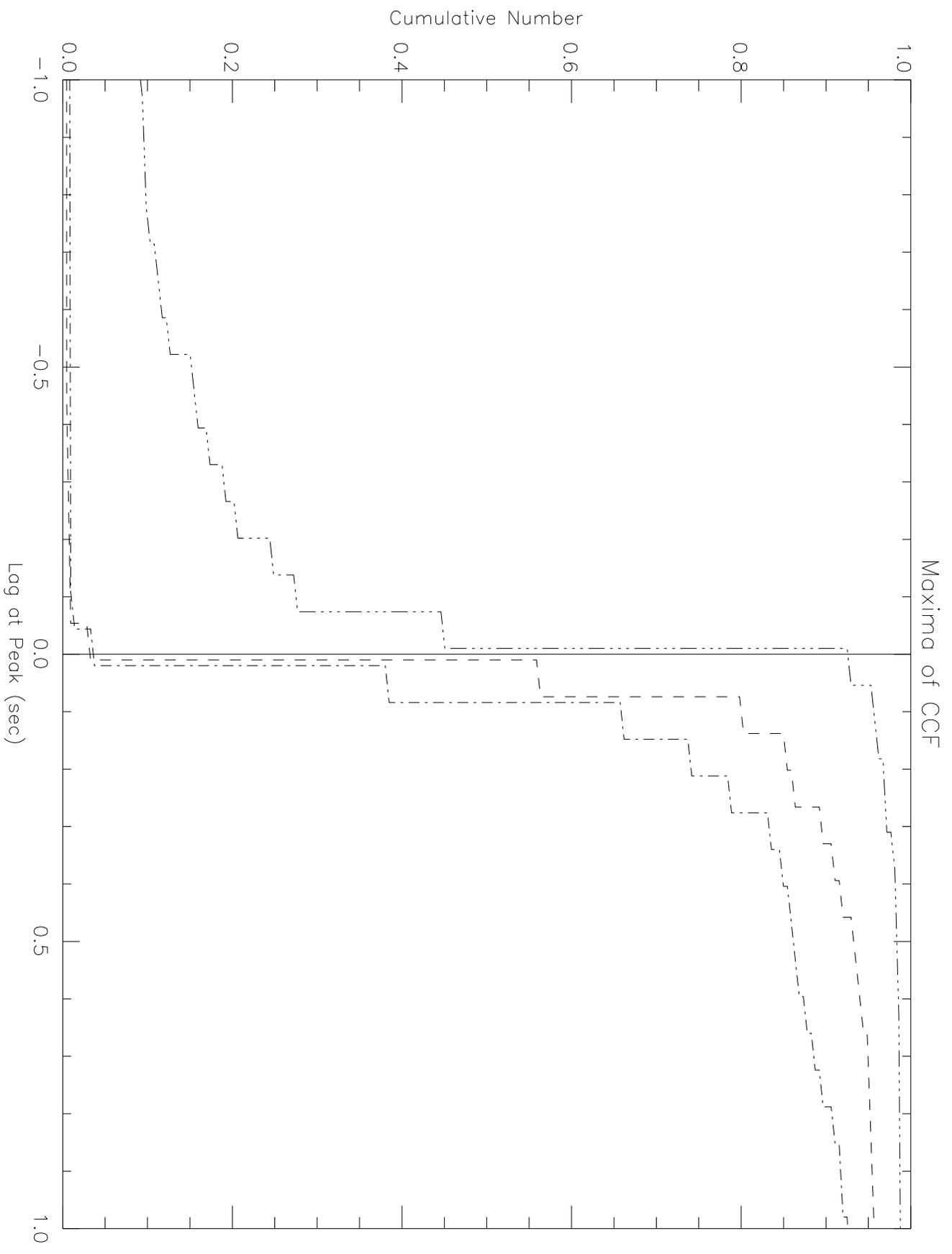
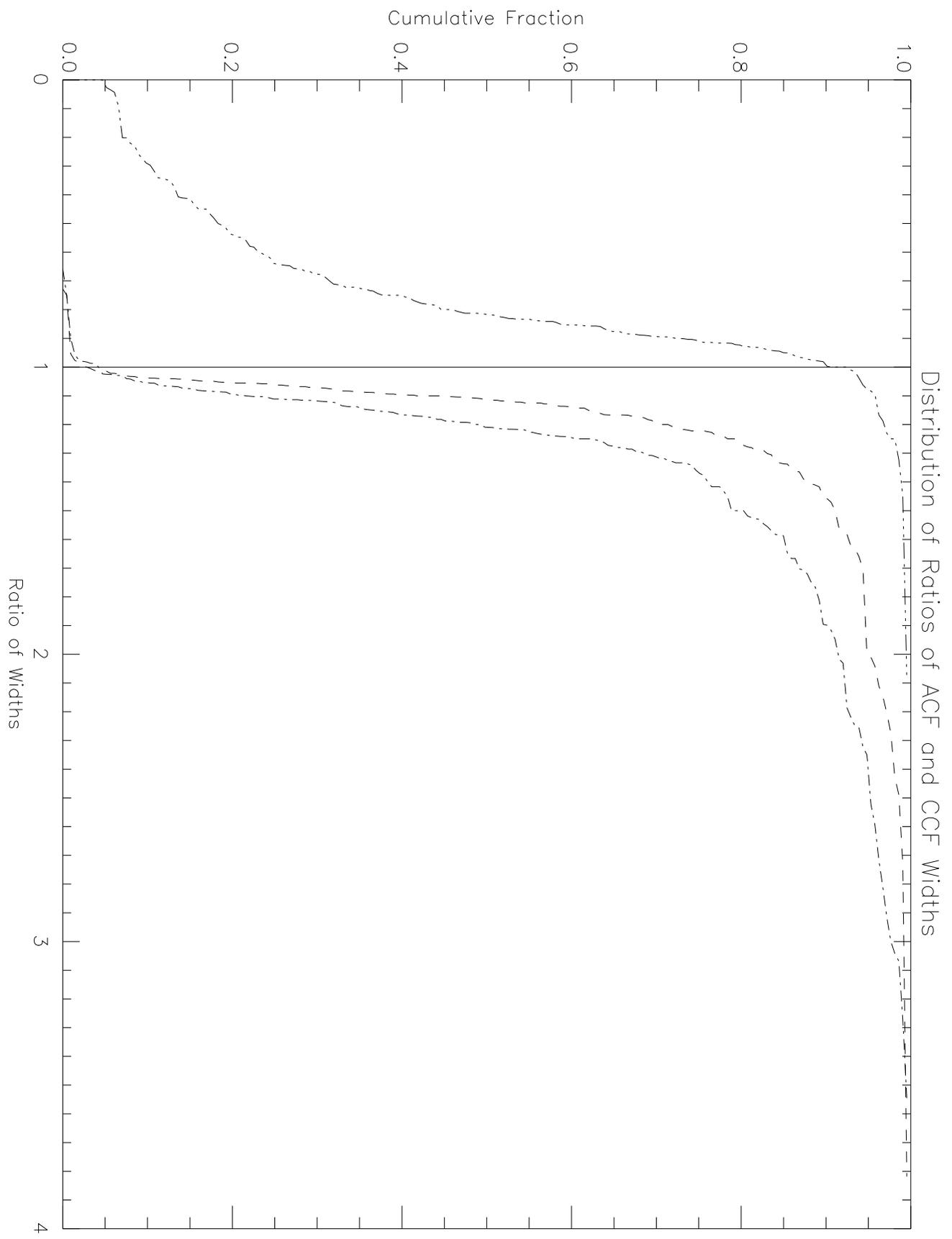
\end{document}